# Conformational Dynamics of a Single Protein Monitored for 24 Hours at Video Rate

Weixiang Ye[1, 2, ‡], Markus Götz[3, ‡], Sirin Celiksoy[1], Laura Tüting[1,2], Christoph Ratzke[3], Janak Prasad[1,2], Julia Ricken[5], Seraphine V. Wegner[5], Rubén Ahijado-Guzmán[1], Thorsten Hugel[3,4*], Carsten Sönnichsen[1*]

[1]Institute of Physical Chemistry, University of Mainz, Duesbergweg 10-14, D-55128 Mainz, Germany

[2]Graduate School Materials Science in Mainz, Staudinger Weg 9, D-55128 Mainz, Germany

[3]Institute of Physical Chemistry, University of Freiburg, Albertstraße 23a, D-79104 Freiburg, Germany

[4]BIOSS Centre for Biological Signaling Studies, University of Freiburg, Germany

[5]Max Planck Institute for Polymer Research, Ackermannweg 10, 55128 Mainz, Germany

‡Contributed equally

*soennichsen@uni-mainz.de; thorsten.hugel@pc.uni-freiburg.de

ABSTRACT: We use plasmon rulers to follow the conformational dynamics of a single protein for up to 24 hours at video rate. The plasmon ruler consists of two gold nanospheres connected by a single protein linker. In our experiment, we follow the dynamics of the molecular chaperon Hsp90 (heat shock protein 90), which is known to show an 'open' and a 'closed' conformation. Our measurements confirm the previously known conformational dynamics with transition times in the second to minute timescale and reveals new dynamics on the timescale of minutes to hours. Plasmon rulers thus extend the observation bandwidth 3-4 orders of magnitude with respect to single-molecule fluorescence resonance energy transfer (FRET) and enable the study of molecular dynamics with unprecedented precision.

The function of proteins is determined by a combination of their structure and dynamics. Acquisition of structural information has recently been revolutionized by advances in electron microscopy,[1,2] however protein dynamics is difficult to obtain over long timescales.[3] Here we show a novel plasmon ruler-based single-molecule approach to study the conformational dynamics of single proteins over 24 hours at video rate in and out of equilibrium (i.e. in the absence and presence of ATP). This approach does not impose external force on the proteins and allows to measure protein dynamics over several orders of magnitude – which cannot be achieved by the established single molecule techniques.[4,5,6] We used this technique to explore the dynamics of the heat shock protein 90 (Hsp90) and find (beside the well-known fast dynamics in the 0.1-10 s range) states with very long dwell times (linked to rarely visited states) in the minute timescale, which could indicate, for instance, pathways to protein denaturation. Our plasmon ruler based method will enable to access the full complexity of correlated local and global conformational dynamics on the level of individual proteins and therefore extend and complement the current 'structure - function' paradigm with a novel 'structure - timescale - function' description. Access to the dynamics of a single protein over more than 6 orders of magnitude further allows us to address important questions like conformational heterogeneity among proteins, ergodic behavior, and non-Markovian dynamics – ultimately extending the often static view of biomolecules with a more dynamic picture.

Light induces the collective oscillation of the conduction electrons in noble metal nanoparticles with a specific resonance frequency called particle plasmon.[7] A plasmon ruler consists of two plasmonic (e.g. gold) nanoparticles bridged by the macromolecule under investigation (**Figure 1a** inset) (see Supplementary Information and **Figures S1** and **S2** for the nanoparticle preparation, characterization and optimization of the assembly). The coupling of the plasmons of the two nanoparticles depends strongly on their separation: decreasing the interparticle distance shifts the plasmon resonance to higher wavelengths ('red shift') with an additional increase of the scattering efficiency (**Figure 1a** and **S3**). By continuously monitoring the scattering intensity of hundreds of single particle pairs (plasmon rulers) in parallel under a dark-field microscope, we determine the interparticle distance and, thus, reveal the dynamics of the bridging macromolecule[8,9] (**Figure 1a** and **S3**). Plasmonic nanoparticles offer the possibility to access a wide range of timescales because of their unlimited photostability and exceptionally strong light scattering. These properties allow us to measure for, in principle, infinite time with microsecond time-resolution (see the Supporting Information for a discussion of the limits of plasmon ruler dynamical measurements, **Figures S4, S5** and **S6** and **Tables S1** and **S2**). Plasmon rulers have been used to study DNA molecules and their interactions with proteins,[10-13] simple polymers[14] and can even be extended to study three dimensional motions.[15] Until now, plasmon rulers have not been used to study conformational dynamics of more complex macromolecules such as proteins. Here, we report key technical and methodological advances in time resolution, nanoparticle functionalization and measurement parallelization that permitted us to follow the dynamics of single proteins for up to 24 hours at video rate (20 Hz) (**Figure 1b** and **S7**). **Figure 1c** shows an exemplary long time-trace from a single molecule. From this one single time-trace, a complete dynamic state model and its transition rates can be deduced. Previously, experiments with single molecule FRET yielded time-traces of around 100 seconds, requiring the pooling of hundreds of single protein traces to extract transition rates. Pooling of traces from different molecules implies ergodicity and the absence of additional dwell times on timescales more than about a factor of two larger than the trace length. This study on the molecular chaperone Hsp90 shows that these assumptions have to be revisited.

We use the Hsp90 as our model system because it has been extensively studied by FRET, X-ray crystallography, and electron microscopy.[16-19] The Hsp90 protein forms homo-dimers, stably connected through the two C-terminal domains and temporarily through the N-terminal domains. The protein complex undergoes scissor-like conformational changes where the N-terminus is 'open' or 'closed'.[16] We sandwiched a single Hsp90 complex between two gold nanospheres in a microfluidic flow cell. In our deposition strategy, one gold nanoparticle serving as anchor was attached to the glass slide. Then, a second particle was attached to the anchor particle with an Hsp90 complex bridging the two as schematically represented in the inset of **Figure 1a**. We use polyethylene glycol (PEG) for the passivation of the flow cell surface and the nanoparticles, as well as molecular linker for the protein-nanoparticle connection (see details and controls in the Supplementary Information **Figures S2, S4, S5** and **S6**). Our assembly strategy results in plasmon rulers linked by a single functioning Hsp90 complex.

Some plasmon rulers in the field of view (usually around 10%) show the telltale signature of the Hsp90 complex known from FRET experiments[16,19]: two distinct conformations (open/closed) with transitions in the second timescale (see **Figure 1c** and **2a-b**). The mean distance change (4.7 ± 0.4 nm) agrees well with the value of 5.2 nm expected from available structures for this mutant (285C-285C), which reads 14.1 nm in the open state[20] and 8.9 nm in the closed state.[21] As additional control, we add the non-hydrolyzable ATP analogue AMP-PNP at the end of the experiment, which is known to lock the protein in the closed state[19]– which we also observe here (**Figure 2c**). All these observations make us confident that the plasmon rulers indeed capture the conformational dynamics of a single Hsp90 dimer. We excluded any influence from the gold nanoparticles on the natural conformational dynamics of Hsp90. First, for 60 nm gold spheres, the diffusion time is around 0.1 μs for a distance change of 1 nm. This diffusion time is around six orders of magnitude faster than the faster Hsp90 dynamics, which is in the range of 0.1 second. We assume that it is possible to measure protein's conformational dynamics slower than this diffusion time. Second, the Hsp90 mutant was selected to avoid any interference or steric hindrance with the gold spheres. Third, the obtained Hsp90 dynamics from the plasmon rulers is in good agreement with the FRET measurements (in the 1 to 100 seconds range).

However, the plasmon ruler traces show dynamics never seen before by FRET: dynamics on the 1-10 min timescale (**Figure 2a**) and even molecules that remain in either the open or closed configuration for several hours (**Figure 2d** and **S7**). Such long-lived states were not observed for plasmon rulers connected only by PEG linkers (**Figure S5**). We compare these plasmon ruler traces (outside of the regions where they remain in one state for extended period of time) to previous FRET studies using an analysis based on a hidden Markov model with four states.[22] **Figure 2e** depicts the four states from the Viterbi path in different colors. In general, the rates extracted from the plasmon rulers agree well with the FRET data (**Figure S8**). What is missing from the FRET data are long lived states. We first quantify states with an intermediate occupancy in the minute timescale. For this analysis, we selected traces (or parts of traces) that are not 'stuck' in one state for hours and converted them into a series of dwell times corresponding to the time spent in the open and closed conformation. In total, we obtained 27700 transitions in the absence of ATP, and 33000 transitions in the presence of ATP (**Figure 3**). The cumulative occurrence $P(t>=\tau)$ data shows linear regions on a negative logarithmic axis for dwell times in the hundreds of second region, which indicates another long-lived sub-state within the open and closed conformation. The lifetimes of these previously unknown states are between four and five minutes for the open conformation and one to two minutes for the closed conformation, with ATP reducing the lifetime in the closed conformation (**Table S3**).

These unexpected long-lived states, especially those on the hour timescale, clearly show the need to observe single molecules for extremely long periods: in this system even 24 hours are not sufficient to ensure ergodicity.[23] In an ergodic system, statistical properties from transitions within a single-molecule trace or from the same amount of transitions from different molecules yield the same result – however, we still find differences (for example of the total time spend in the open configuration) between 24 hour traces (**Figures S7** and **S9**). Ergodicity is an assumption generally taken in the single-molecule field,[24] which needs to be revisited in the light of these results. The long-lived states could correspond to misfolded protein domains. This would fit well with the findings from optical tweezer experiments that domain unfolding at zero force occurs on the timescale of hours and that the generally very fast refolding is often hindered by long lived misfolded states for the full length multi domain Hsp90 protein.[25,26]

The analysis of these transitions within the Hsp90 molecule shows the capability of plasmon rulers to find conformational dynamics at both fast and slow timescales from one single protein complex. The accessible bandwidth is paralleled only by electrophysiological measurements on single ion channels with the patch-clamp technique, which led groundbreaking discoveries in the 1990s.[27] Not only the bandwidth is remarkable, also the signal to noise level is at least a factor of four better at comparable time resolution than single molecule FRET (**Figure S10**). The versatility and simplicity of plasmon rulers as tool to study single protein conformational dynamics makes it possible to study slow and rare processes such as protein misfolding and denaturation, or the complex conformational changes after or during the interaction with other proteins or small molecules. Comparing single-molecule traces will show the dynamical effects of the variability (non-ergodicity) of nominally identical protein species, caused for example by small differences in the secondary or tertiary structure or post-translational modifications. More fundamentally, long single molecule traces open the window to investigate memory effects (non-Markovian behavior) and directionality (detailed balance).[23]

ASSOCIATED CONTENT

**Supplementary Information** containing protein purification and labelling, nanoparticle synthesis, functionalization, flow cell preparation, plasmon ruler formation, details on the setup and setup performance, principal limits of the plasmon rulers study, decay rates and comparison with FRET is available.

AUTHOR INFORMATION

The authors declare no competing financial interests.

**Corresponding Author**

*thorsten.hugel@pc.uni-freiburg.de

*soennichsen@uni-mainz.de

**Author Contributions.** C.S. and T.H. initiated and designed the research. The Hsp90 mutants were expressed and purified by M.G. and C.R.. The measurement setup and analysis software was developed by S.C. and W.Y. under the guidance of C.S., the dimer formation protocol was initially developed by J.P., then modified by L.T., both under the guidance of R.A-G. and C.S. The single molecule traces reported in this manuscript were measured by W.Y., L.T., S.C., M.G., R.A-G. The interpretation of the dimer traces and their statistical analysis was performed by W.Y., L.T., S.C., M.G., R.A-G, T.H. and C.S. The manuscript text was written through contributions of all authors. All authors have given approval to the final version of the manuscript.

ACKNOWLEDGMENT

This work was financially supported by the ERC grants 259640 (SingleSens) and 681891 (Prosint). L.T and W.Y. are recipients of DFG fellowships through the Excellence Initiative by the Graduate School Materials Science in Mainz (GSC 266). We thank Björn Reinhard (T.H.) and Friederike Schmid (C.S.) for helpful discussions. We thank Karl Wandner and Eva Wächtersbach for technical assistance. Karl Wandner made significant contributions to the setup control and data analysis software. Sebastian Schmachtel contributed to the dimer formation protocols, Arpad Jakab contributed significantly to the development of one of the setups. Jana Strugatchi and Johannes W. Sutter participated in some of the single molecule measurements under the guidance of S.C. and W.Y.

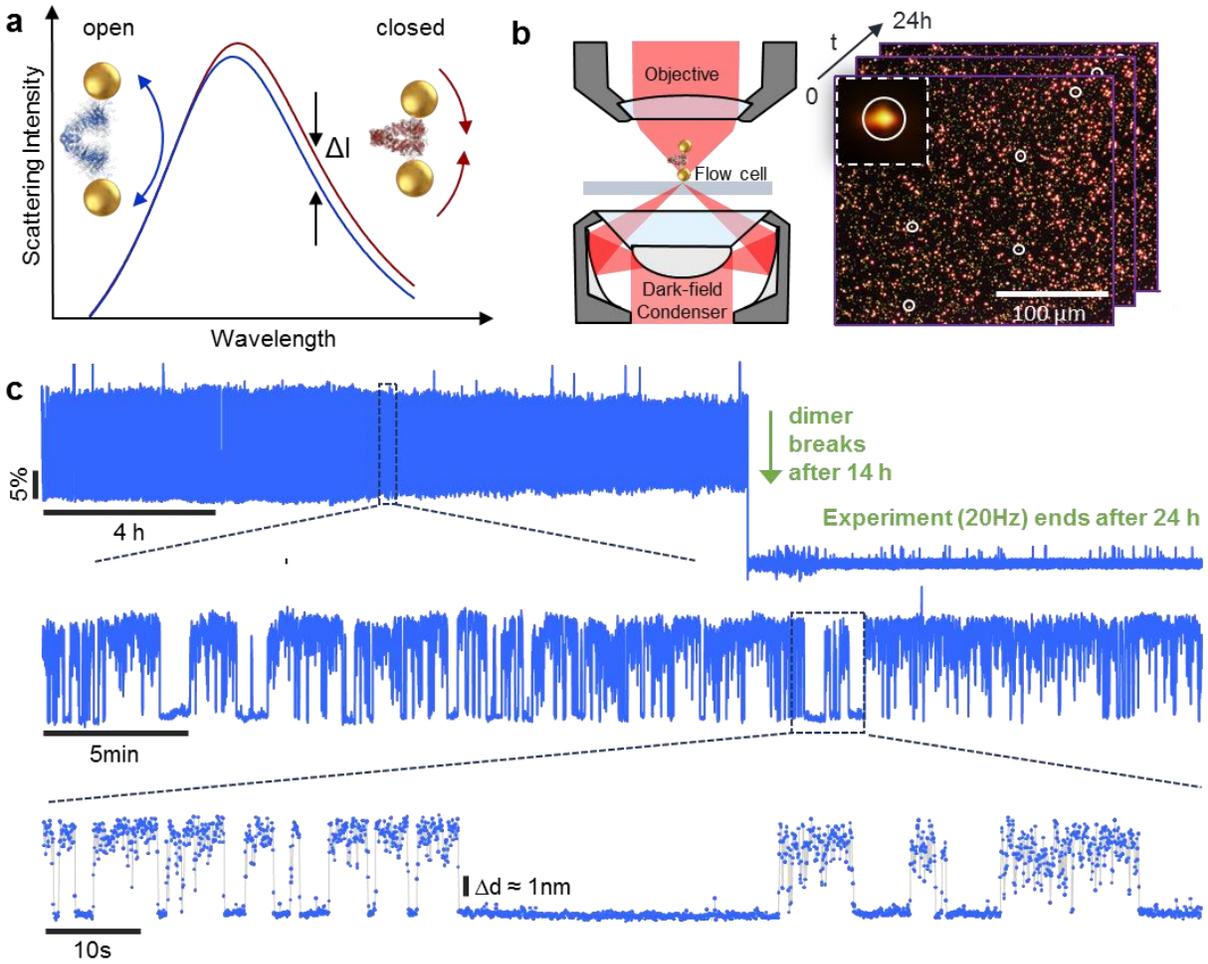

**Figure 1. Plasmon rulers show Hsp90 dynamics on millisecond to hours timescale. a**. Change of interparticle distance leads to shift in scattering spectra and thus allows to distinguish the open (blue line) and closed (red) configuration of an Hsp90 sandwiched between two plasmonic nanoparticles. Both, overall intensity and the intensity at a given wavelength (black arrows) changes with interparticle distance. **b**. Many Hsp90 linked plasmon rulers can be observed in parallel. Under a dark-field microscope (left side), the scattered light from the nanoparticles can be collected with low background (right side). We follow the intensity of those dots over time (examples indicated by the white circles). **c**. The relative intensity (normalized to its mean) of a single plasmon ruler (blue line) as a function of time. In this example, the plasmon ruler was measured every 50ms (at 20Hz) for 24 hours. After about 14 hours, the dimer breaks in a single step (confirming that only one linker connected the dimer initially). Dynamics can be observed at timescales ranging from hours, over minutes towards milliseconds. We show a zoom into 40 min of the fluctuating part and another zoom into 2 min (indicated by the black dashed line). The zoom at the bottom shows individual data points (blue circles), connected by gray lines as a guide to the eye.

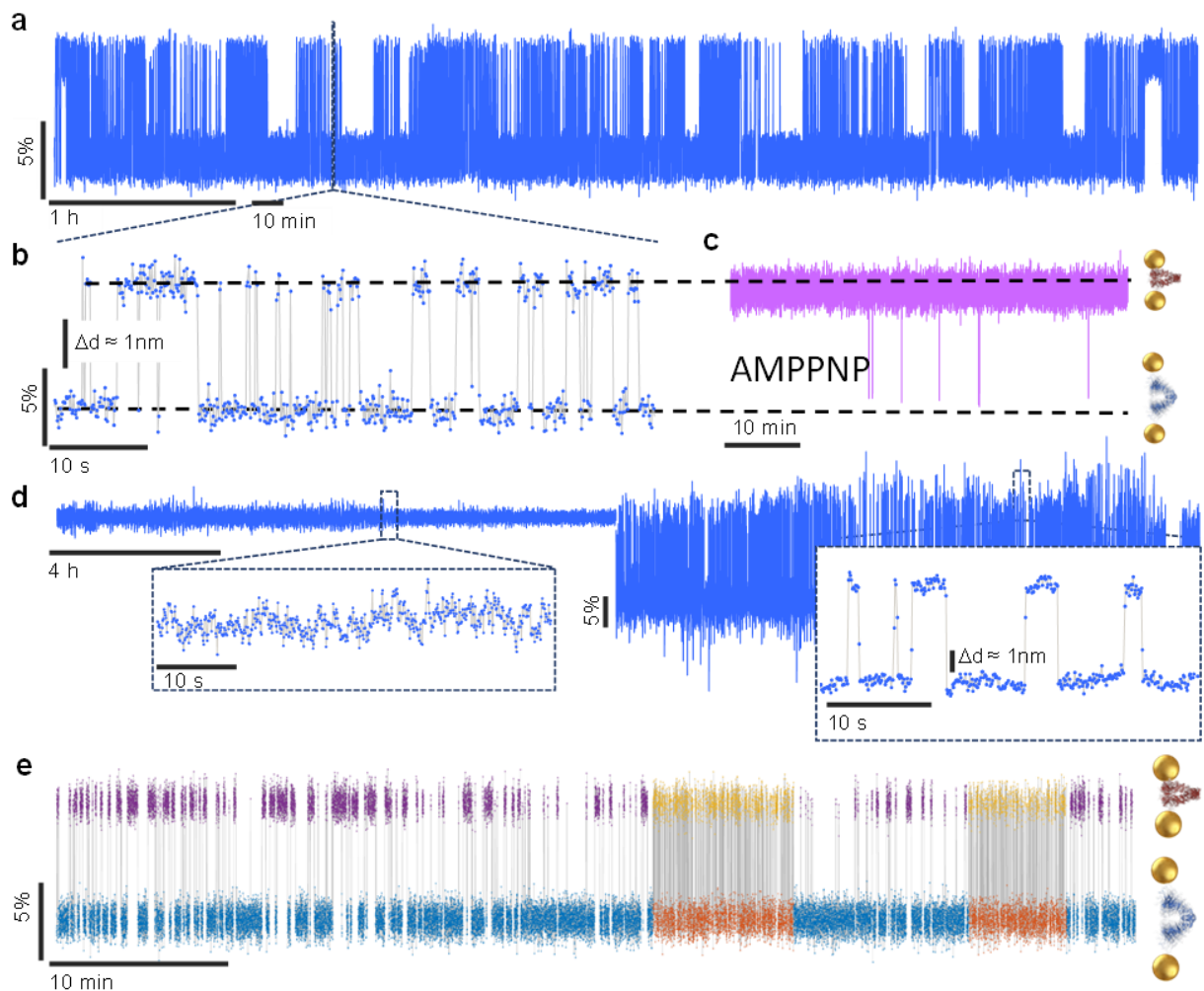

**Figure 2. Nucleotide dependent single molecule time-traces of Hsp90 .a.** Example of a timetrace of Hsp90 transitions between open and closed states recorded for 6 hours at a time resolution of 100 ms. **b.** Zoom into the timetrace shown in 'a' as indicated by the dashed lines. The blue dots are the actual data points, connected by gray lines as a guide to the eye. The open and closed state are clearly separated in relative intensity. The relative intensity can be roughly converted to separation distance Δd in nm as indicated by the second vertical scale bar. **c.** The same plasmon ruler is locked in its closed state for many minutes after addition of AMP-PNP to the buffer (pink line), which indicates a functional protein. **d.** Another example of an Hsp90 timetrace, recorded for 24 hours at 20 Hz. This example shows the situation where Hsp90 is 'stuck' in its closed conformation for the initial 12 hours before resuming rapid transitions between the open and closed state. Insets show zoom-in of both parts. **e**. Example of a part of a timetrace where the data points are colored according to the most likely sub-states determined by a Hidden Markov analysis with four states.

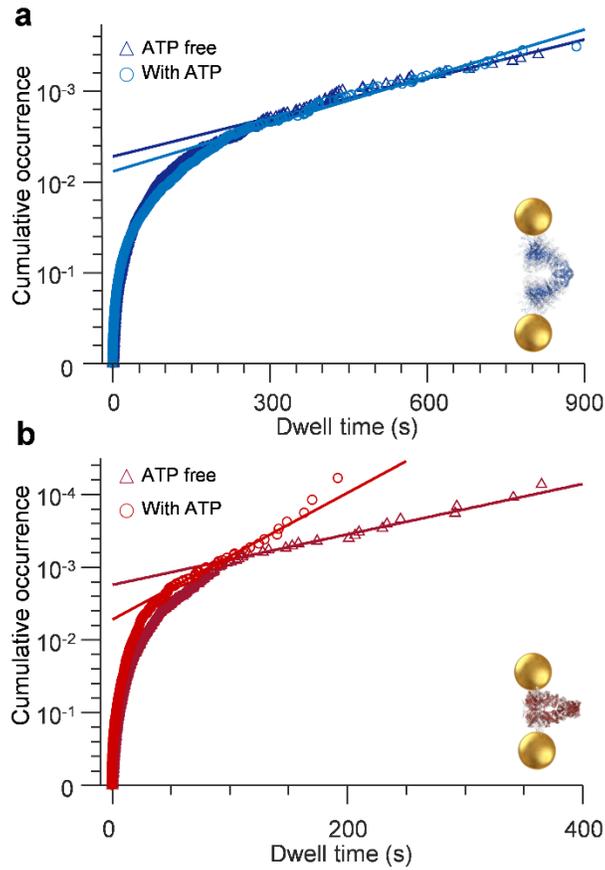

**Figure 3. Quantification of previously inaccessible slow dynamics:** The cumulative dwell time distribution $P(\tau)$ gives the probability to find dwell times shorter than $\tau$. To make rare states with long lifetimes more visible, we display the cumulative occurrence $P(t>=\tau)$ on a (negative) logarithmic axis, where a statistically independent process is represented as a straight line. The triangles correspond to an ATP free buffer, the circles to an experiment in the presence of ATP. **a** shows the cumulative occurrence for the open configuration, **b** for the closed configuration. In both cases, there is a linear region with slopes corresponding to dwell times in the 100 s of second regime (blue and red lines). This long-lived state is significantly affected by ATP in the closed configuration.

TOC Figure

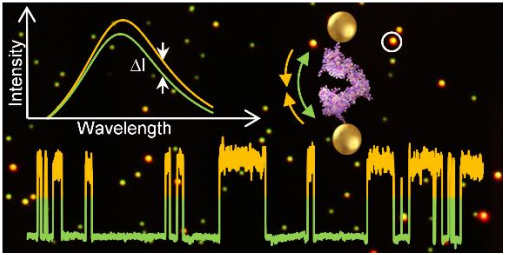

Supplementary Information for

# Conformational Dynamics of a Single Protein Monitored for 24 Hours at Video Rate

*Weixiang Ye*[1,2,‡], *Markus Götz*[3,‡], *Sirin Celiksoy*[1], *Laura Tüting*[1,2], *Christoph Ratzke*[3], *Janak Prasad*[1,2], *Julia Ricken*[5], *Seraphine V. Wegner*[5], *Rubén Ahijado-Guzmán*[1], *Thorsten Hugel*[3,4*], *Carsten Sönnichsen*[1*]

[1]Institute of Physical Chemistry, University of Mainz, Duesbergweg 10-14, D-55128 Mainz, Germany

[2]Graduate School Materials Science in Mainz, Staudinger Weg 9, D-55128 Mainz, Germany

[3]Institute of Physical Chemistry, University of Freiburg, Albertstraße 23a, D-79104 Freiburg, Germany

[4]BIOSS Centre for Biological Signaling Studies, University of Freiburg, Germany

[5]Max Planck Institute for Polymer Research, Ackermannweg 10, 55128 Mainz, Germany

**Contents**

# Methods

## Materials

Chemicals were acquired from Sigma-Aldrich or Merck in analytical grade. We used deionized water from a Millipore system (> 18 MΩ, Milli Q).

## Protein purification

Hsp90 (Hsp82 in yeast, UniProtKB ID P02829) was expressed as a fusion protein with a C-terminal coiled-coil motif from the kinesin neck region of *Drosophila melanogaster* (DmKHC) that prevents dimer dissociation at picomolar concentrations in single molecule experiments[1]. Wild-type yeast Hsp90, containing no cysteines, was modified by site-directed mutagenesis (QuikChange Lightning, Agilent, Santa Clara, California) to attach exactly one fluorescent dye or gold nanoparticle per subunit of the Hsp90 dimer. For the plasmon ruler experiments, the T285C mutant with a non-cleavable N-terminal His-tag and a Strep-tag at the far C-terminus was used. Single molecule FRET (smFRET) experiments were conducted with a N298C single cysteine variant which additionally carries a tag for *in vivo* biotinylation (AviTag) at the far C-terminus. The N-terminal His-SUMO-tag is cleaved during purification.

The Hsp90 constructs are contained in pET28 vectors and expressed from *Escherichia coli* BL21 (DE3) or BL21 Star (DE3) cells in the case of AviTag constructs (Thermo Fisher Scientific, Waltham, Massachusetts).

Standard expression was done in LB medium supplemented with 50 μg/mL kanamycin at 37°C and inoculation from an over-night culture (1:100). Cells were induced at an OD600 of 0.6 by addition of 1 mM IPTG. After 3 hours, the cells were harvested by centrifugation (20 min, 4°C, 3'000 rpm, JLA 8.1, Avanti JXN-26, Beckman Coulter), resuspended in phosphate buffered saline (PBS) and pelleted again (10 min, 4°C, 4700 rcf Rotanta 460R, Hettich, Tuttlingen, Germany).

For *in vivo* biotinylation, the biotin ligase (BirA) is co-expressed from pBirAcm (Avidity LLC, Aurora, Colorado) following the manufacturer's instructions. Briefly, TB medium was supplemented with 0.5% glucose and 30 μg/mL kanamycin. Expression was induced at an OD600 of 0.7 by adding 1 mM IPTG and 50 μM d-biotin (from a 5 mM stock, in warm 10 mM bicine buffer pH 8.3, filter-sterilized). After induction for 3 h at 37°C, cells were harvested by centrifugation, washed with PBS and pelleted again.

For purification, the cells were re-suspended in approximately 30mL PBS and lysed with a Cell Disruptor (Constant Systems) at 1.6 kbar. Cell debris was pelleted by centrifugation at 30'000$g$ at 4°C for 45 min (JA-25.50, Avanti JXN-26, Beckman Coulter) and the supernatant was cleared by additional filtration (Filtropur S 0.45, Sarstedt, Nümbrecht, Germany).

To prepare the protein for the plasmon ruler experiment, the solution was applied to a gravity flow Strep-Tactin column (IBA GmbH, Göttingen) and eluted according to the manufacture's protocol.

For the smFRET construct, 20 mM imidazole was added from a 1 M stock, the solution was applied to a 5 mL HisTrap HP (GE Healthcare, Freiburg, Germany) and eluted by a linear gradient from 20 to 500 mM imidazole in 50 mM sodium phosphate pH 8.0, 300 mM NaCl at 8°C. Protein-containing fractions were pooled and dialyzed against the imidazole-free buffer overnight in the presence of 1/100 equivalent SENP protease. This protease cuts off the N-terminal His-SUMO sequence, leaving the native, tag-free protein[2]. The solution is again applied to the HisTrap column and the flow-through is collected and diluted 1:3 with MilliQ water to decrease the ionic strength.

Both constructs were then applied to a HiTrap Q HP 5mL (GE Healthcare) and the protein was eluted with a linear gradient from 50 mM to 1 M NaCl in 40 mM HEPES pH 7.5. Hsp90 fractions were pooled and concentrated using centrifugal filters with a 50 kDa molecular weight cut-off (Amicon Ultra, Merck Millipore, Darmstadt, Germany). Finally, the protein was applied to a gel filtration column (HiLoad 16/600 Superdex200, GE Healthcare) and eluted with 40 mM HEPES, 200 mM KCl pH 7.5. Peak fractions were again pooled and concentrated to 50-100 μM.

## Fluorescence labelling and monomer exchange

Typically, 50 μL of a 50 μM Hsp90 solution were reduced for 30 min at room temperature with 5–10 mM TCEP, added from a pH-adjusted stock solution. Buffer was exchanged to 1× PBS, pH 6.7 using an Amicon Ultra 0.5mL centrifugal filter (50 kDa MWCO, Merck Millipore) and an appropriate dilution factor. The sample were brought to a concentration range of 30–70 μM after the last centrifugation step. Maleimide derivatives of the dyes (Atto 550 or 647N, ATTO-TEC, Siegen, Germany) were added from a millimolar stock solution in DMSO in 1.2–1.5× excess and allowed to react for 60 min at room temperature in the dark. Free dye was removed using PD MiniTrap G-25 columns (GE Healthcare) with the spin protocol and smFRET measurement buffer (40 mM HEPES, 150 mM KCl, 10 mM $MgCl_2$, pH 7.5). Potential aggregates were removed by centrifugation (4°C, 14'000 rpm, >30 min) and aliquots were snap-frozen in liquid nitrogen and stored at -80°C.

To form hetero-dimers with one donor and one acceptor label, 250 nM of each labelled species was incubated at 47°C for 30 min in smFRET measurement buffer. This allows exchanging of the Hsp90 subunits by destabilizing the C-terminal coiled-coil motif. Aggregates were removed by subsequent centrifugation in a benchtop centrifuge at 4°C for 60 min.

## Nanoparticle synthesis

The gold nanoparticle size to choose depends on the range of interparticle distances to study. Both, particle diameter and interparticle distance determine the plasmon coupling strength. This dependency was well described by a single exponential decay with a decay length around 1/4 of the particle diameter.[3,4] Following this rule we chose 62.5 nm gold spheres as optimum for Hsp90.. The gold nanospheres used in this work were prepared as described in references 3 and 4.[5,6] The particle sizes were determined by transmission electron microscopy (TEM). **Figure S1** shows representative TEM image. From the image, we determined the mean and standard deviation of diameter (D=62.5±2.50 nm) for 127 gold nanospheres. These particles show single-particle plasmon resonance wavelength of 588 ± 8 nm when immobilized on a glass substrate (in buffer).

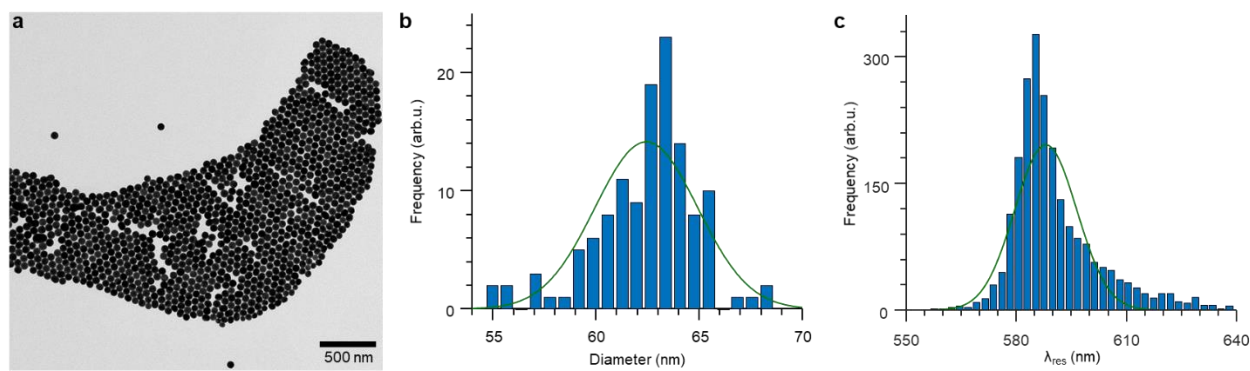

**Figure S1 | Characterization of gold nanospheres. a**, Representative transmission electron microscopy image of the gold nanospheres used in this work. **b**, The diameter distribution of the nanospheres from 127 individual particle is 62.5±2.5 nm.**c**, The resonance wavelength distribution of 2000 individual nanoparticles is 588 ± 8 nm.

## Nanoparticle functionalization

**AuNP@PEG-Biotin.** 500 µL AuNPs were centrifuged (5000 $g$, 5 min), the supernatant was removed and the pelleted nanoparticles were resuspended in 100 µL freshly prepared 2 mM PEG mixture (thiol-PEG-OMe 2 kDa, thiol-PEG-biotin 3 kDa, thiol-PEG-NH$_2$ 3 kDa (Iris Biotech GmbH) with a molar ratio of 87:10:3 and incubated for 24 hours under stirring at room temperature. To remove the excess of unbound PEG, we washed the nanoparticles (AuNP@PEG-Biotin) twice by centrifugation (5000 $g$, 5 min) with 400 µL of Milli-Q water and stored at 4°C until use.

**AuNP@PEG-NH$_2$.** 500 µL AuNPs were centrifuged (5000 $g$, 5 min), the supernatant was removed and the pelleted nanoparticles were resuspended in 100 µL freshly prepared 2 mM PEG mixture (thiol-PEG-OMe 2 kDa, thiol-PEG-NH$_2$ 3 kDa with a molar ratio of 97:3, Iris Biotech GmbH).The mixture was incubated under stirring for 24 hours at room temperature. To remove the excess of unbound PEG, we washed the nanoparticles twice by centrifugation (5000 $g$, 5 min) with 400 µL of Milli-Q water and stored at 4°C until use.

## Flow cell preparation

The glass slides used to prepare our flow cell were pegylated and biotinylated as described in reference 5.[7] After the assembly of the flow cell, we incubated with a streptavidin solution (250 µg/mL in PBS, pH=7.4) for 15 min and then washed with PBS to remove unbound streptavidin. Thus, the inner surface of our flow cell is covered with streptavidin for the later attachment of the particles as explained below.

## Dimer formation

We incubated a diluted AuNP@PEG-Biotin solution (in PBS) in our flow cell for 5 minutes. After the immobilization of approximately 1000 nanoparticles in our field of view, we rinsed our flow cell with PBS to wash away unbound nanoparticles (**Figure S2a,b**). Then, we incubated the immobilized nanoparticles with 1 mM SMCC (succinimidyl 4-(N-maleimidomethyl) cyclohexane-1-carboxylate) in PBS pH=7.4 for 30 min. After a washing of the flow cell with PBS pH=6.7 for 5 minutes, we flushed in our flow cell a solution containing 0.24 nM Hsp90 and 200 nM TCEP in PBS pH=6.7. After an incubation time of 30 minutes, the flow cell was washed again with PBS pH=6.7. Meanwhile, the second nanoparticle stock (AuNP@PEG-NH$_2$) was incubated for 20 min with 1 mM SMCC, and washed by centrifugation four times with water. After the last centrifugation, the pelleted nanoparticles were resuspended in PBS pH=6.7, inserted into the flow cell and incubated for 30 min for the single-molecule plasmon ruler formation. Unbound particles were removed by HEPES buffer (40 mM HEPES, 50 mM KCl, 10 mM MgCl$_2$, pH=7.4) (**Figure S2a,b**).

## Optimization of the dimer assembly process

To study and optimize the dimer formation in the flow cell, we varied the salt concentration of the buffer (**Figure S2c**) and the protein concentration (**Figure S2d**). We found that by using 50 mM KCl and 10 mM MgCl$_2$ and for a protein concentration of 0.24 nM our conditions improved. **Figure S2e** shows the ratio between monomers, non-functional dimers and functional dimers.

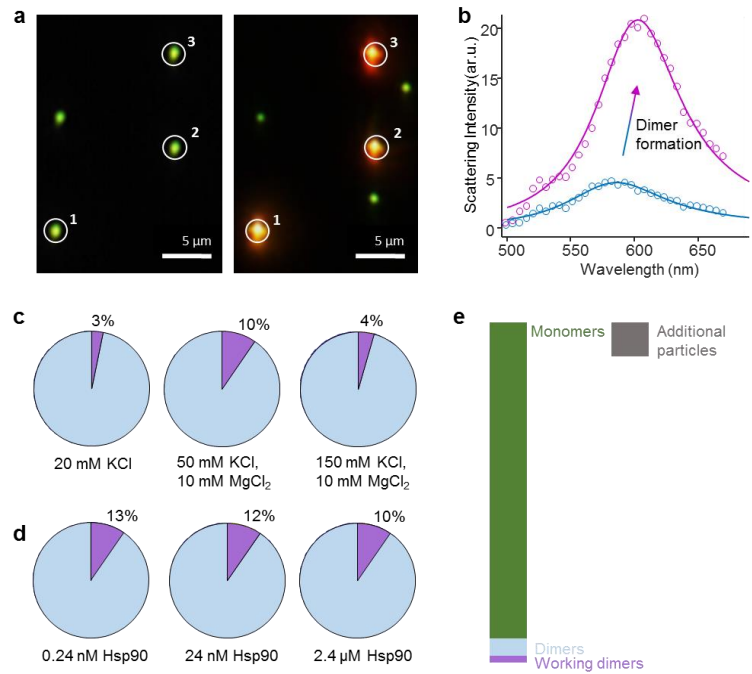

**Figure S2 | Optimization of dimer (plasmon ruler) assembly process. a**, Real color image of gold nanospheres in a dark-field microscope. Individual nanospheres (monomers) appear as green dots (left), some become orange after a second particle is bound (right side). Software picks up such cases (indicated by numbered circles) .**b**, Scattering spectra of these monomers and dimers, showing an increase of overall intensity as well as a red-shift of the plasmon maximum. **c**, Percentage of dimers fluctuating (violet). In lower salt concentration unspecific attachment of particles is hindered, but protein denaturation is more likely compared to higher salt concentrations. We observe the optimum salt concentration to be 50 mM KCl. **d**, Variation of Hsp90 concentration shows no significant change detectable. To decrease the probability for double-linker formation, we choose to use 0.24 nM Hsp90. **e**, Ratio between monomers, non-functional dimers and functional dimers. The 'additional particles' are small nanoparticle aggregates or impurities and defects of the flowcell.

## Measurement conditions

We used for all the experiments the HEPES buffer containing 40 mM HEPES, 50 mM KCl, 10 mM MgCl$_2$, at pH=7.4 supplemented with/without ATP as specified. After carrying out the experiment (with (5 mM) or without ATP) and observing the thermally-induced fluctuations, we incubated our plasmon rulers with the same buffer supplemented with 2 mM AMP-PNP. The non-hydrolizable ATP analog AMP-PNP is known to lock the Hsp90 complex in the closed state. After the complete exchange of the buffer in the flow cell, we started to record a new timetrace.

## Microscope setup

The dark-field setups used Plan–Apochromat 40x/1.3 Zeiss objectives and a SuperK EXTREME supercontinuum laser with a SuperK SELECT multi–line tunable filter (NKT Photonics) or fixed wavelength LED (623nm) as light source. For the automated data acquisition (Hamamatsu orca flash V4.0) and data analysis, we used a MATLAB−based software. After the data acquisition, we corrected the baseline of our time traces by low-pass filtering, and normalized the signal to its mean.

## Conversion of intensity to distance

We estimate the equilibrium interparticle distance $d_0$ for our Hsp90 plasmon ruler (PEG - Hsp90 - PEG) to be approximately $d_0 = 19$ nm (16.9 nm in the closed state and 22 nm in the open using the known dimensions of Hsp90 (8.9 nm - 14 nm) and PEG linkers (4 nm)). By using Boundary Element Method[8] (BEM), we estimated the spectral response of our plasmon rulers to changes in the interparticle distance (**Figure S3a**). The scattering intensity $I_{rel}(d)$ at a given wavelength $\lambda_0$ (normalized to the intensity measured at the estimated equilibrium particle separation $d_0 = 19$ nm) decreases with increasing interparticle separation $d$ (**Figure S3b**). Around the equilibrium particle separation $d_0$, the function $I_{rel}(d)$ shows a nearly linear relationship (purple line **Figure S3b (inset)**). Within the maximum range of expected interparticle distances (about 15-23nm), this linear approximation works well (**Figure S3c (inset)**). We use the slope of this line to convert relative intensity changes $\Delta I_{rel}$ to interparticle distance changes $\Delta d$ ($\Delta d = w \cdot \Delta I_{rel}$). For three different wavelength $\lambda_0$ (615nm, 625nm, 635nm), we extract $w = 27.6$, $w = 24.5$ and $w = 24.3$, respectively. Given all of the uncertainties in this conversion (PEG linker length, exact sizes of the nanoparticles in a given dimer, etc.), we believe that $w = 25$ is a reasonable conversion factor to convert relative intensity to distance changes. We like to stress that it is much easier to interpret the time evolution of the (relative) signal intensities compared to the absolute values for the measured distances, which have a large systematic error, similar to most other single molecule techniques.

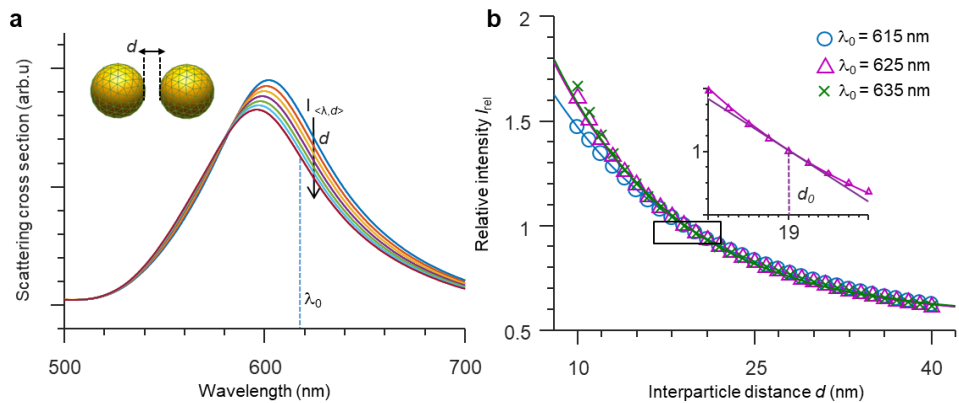

**Figure S3 | Conversion of scattering intensity to interparticle distance. a**, Simulated scattering spectra for different interparticle distance $d$. In the simulations, the gold spheres had a diameter of 62 nm and were surrounded by a medium with a refractive index $n = 1.34$. We used tabulated optical constants for gold. **b**, Relative intensity changes at a fixed wavelength $\lambda_0$ as a function of interparticle distance $d$. At the equilibrium distance of our plasmon rulers ($d_0 = 19$ nm), the relative intensity depends approximately linear on d with a slope of $w = 25$ (inset).

# Setup performance and principal limits for plasmon rulers

The measured fluctuations of the signal obtained from a single plasmon ruler are caused both by measurement noise (1) and by real interparticle distance changes (2):

1. The light sources, optics and detectors lead to a fluctuating signal level, which we refer to as **'setup noise'**. The fundamental reason for this noise is the statistical nature of light which is made from a discrete number of photons. On top of this 'photon shot noise', there is noise in the detector and it is analog to digital converter. Also the light emission and collection efficiency might not be completely stable over time, for example by small changes in the focus position ('drift').
2. When we measure plasmon rulers connected by Hsp90, we observe an increased fluctuation amplitude compared to the above mentioned setup noise. These fluctuations are 'real' in the sense that they are caused by changes in the interparticle distance $d$. Part of this additional fluctuation is due to the PEG spacers, the **'linker contribution'**, another part is due to the stiffness of the Hsp90 linker itself, the **'Hsp90 contribution'**. On top of both of these additional fluctuation sources, the Hsp90 shows the distinct transitions between the open and closed conformation – the topic of this work. In this context, the above mentioned linker and Hsp90 contributions are 'noise', even though part of it contains information about molecular dynamics.

We characterized the contribution of both the setup noise of our current setups (which is, of course, dependent on exposure time and the amount of available light) and the contribution from the linker ('Linker contribution') and the Hsp90 molecule itself. As a measure of the different contribution to the fluctuating signal, we use here the standard deviation of the signal over a given period of time (10 min). We estimate these contributions on several long (24 hours) timetraces and report the average of the obtained values and their standard deviation.

If we try to estimate the resolution of the plasmon rulers for detecting conformational changes of macromolecules, we need also to consider the time-resolution and bandwidth of the measurement: On the other hand, longer measurements have additional sources of noise ('drift'), on the other hand, reducing the time-resolution (increasing exposure time) decreases all statistical noise. In the section '**distance resolution**' we give a few examples of the ability of plasmon rulers to report distance changes on various timescales. At the end, we discuss in the section **'limits of plasmon rulers'** what causes noise on a fundamental level and how much the distance resolution could be improved.

## Setup noise

To estimate the setup noise, we measured gold nanorods with approximately the same scattering intensity as our plasmon rulers (**Figure S4**). These nanorods consist of a solid rod-shaped nanoparticle which prohibits any fluctuations coming from interparticle distance variations. The remaining signal fluctuations are therefore caused by the measurement process itself, probably because of noise in the camera and the light source. These rod-shaped particles have a fluctuation amplitude of around 0.28% ± 0.10% over 10min. Reducing the exposure time to 1s / 10s (by averaging 10 / 100 data points), the fluctuation amplitude reduces to 0.21% ± 0.10 % / 0.17% ± 0.08 %, respectively.

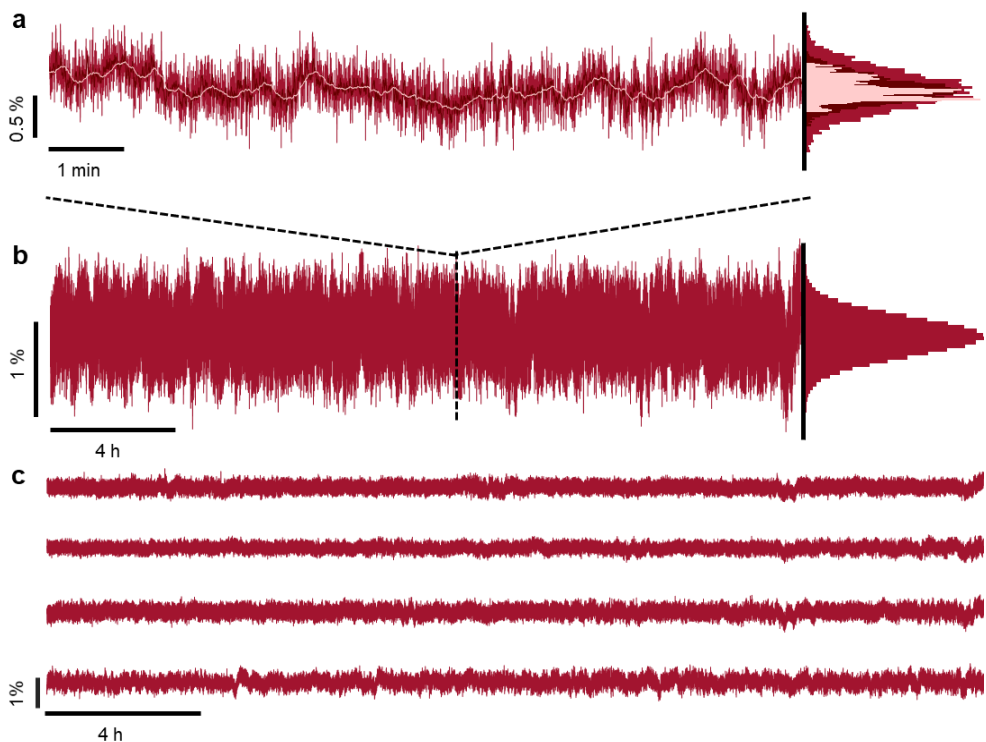

**Figure S4 | Time traces of nanoparticles without (soft) linker.** These time traces were measured on particles that are not connected by soft linkers. On these particles, nothing should be changing over time, so that the signal variation is caused only by noise in the measurement itself. We show a zoom of 10 min in **a**, the entire 24 hours measurement in **b**. **c**, Additional traces of different gold nanorods. We evaluate the signal variation over 10 min. with the full time resolution (100ms, red line), with 1s time resolution (medium light red), and 10s time resolution (light red), c.f. histograms in **a**, right side. The observed signal variation of 0.4% is caused by setup noise.

## Linker contribution

To estimate the linker contribution, we substituted the Hsp90 protein dimer by a dithiol-Poly(ethylene glycol), 5 kDa, Rapp Polymere GmbH. **Figure S5** shows some examples of such PEG linked plasmon rulers measured with 100ms time-resolution. These PEG dimers have a fluctuation amplitude of around 0.80% ± 0.62 % over 10min. Reducing the exposure time to 1s / 10s (by averaging 10 / 100 data points), the fluctuation amplitude reduces to 0.64% ± 0.58 % / 0.53% ± 0.49 %, respectively.

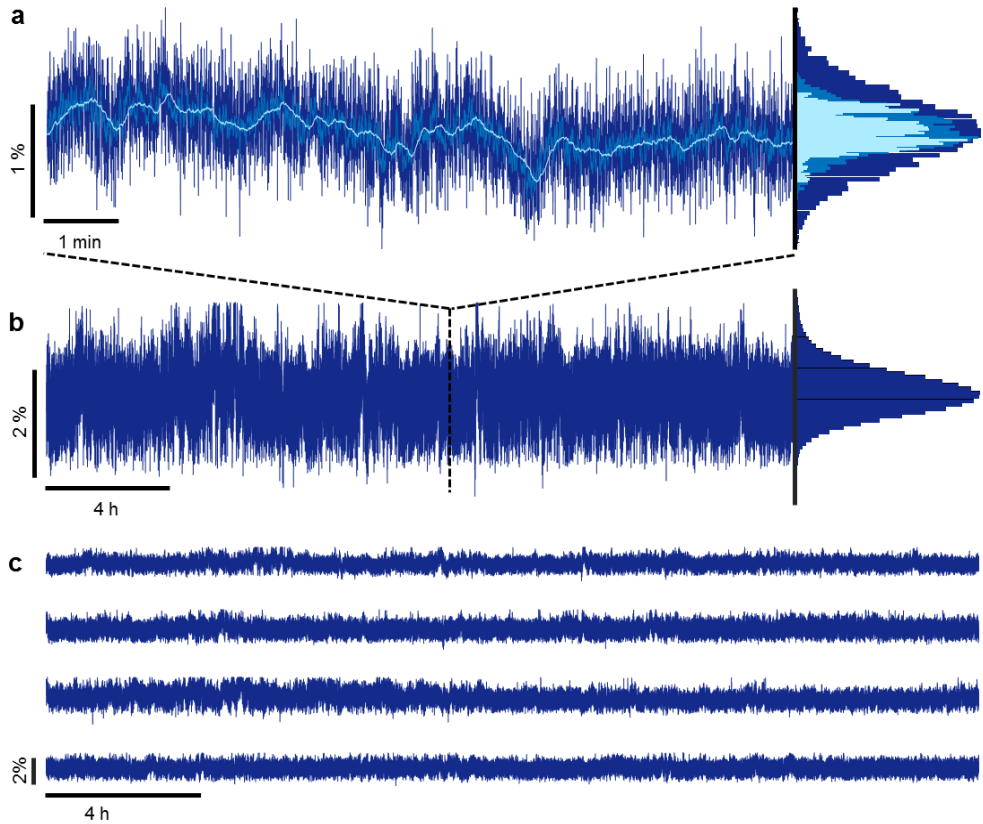

**Figure S5 | Time traces of PEG connected plasmon rulers.** Time traces of plasmon rulers connected by PEG molecules, zoom of 10 min in **a**, the entire 24 hours measurement in **b**. **c**, Additional traces of different plasmon rulers. We evaluate the signal variation over 10 min. with the full time resolution (100ms, blue line), with 1s time resolution (medium light blue), and 10s time resolution (light blue), c.f. histograms in **a**, right side. The observed signal variation of 0.8% is caused, in part, by the PEG linkers.

## Hsp90 contribution

Plasmon rulers connected by Hsp90 show a larger fluctuation amplitude than PEG connected dimers, presumably due to the lower rigidity of Hsp90 compared to PEG (in the given buffer conditions), see **Figure S6**. This Hsp90 contribution is not caused by the opening or closing of the Hsp90 dimers, which show much larger and clearly distinct fluctuations. It is not clear to us why some (or most) of the Hsp90 linkers do not show the opening/closing behavior. In some cases, there could be multiple linkers between the two nanoparticles but it is unlikely that this is the only reason for this behavior. It is possible that our plasmon ruler preparation procedure denatures many Hsp90 molecules in a way that make them unable to show the opening/closing transition.

These Hsp90 dimers have a fluctuation amplitude of around 1.98% ± 0.87 % over 10 min. Reducing the exposure time to 1s / 10s (by averaging 10 / 100 data points), the fluctuation amplitude reduces to 1.85% ± 0.89 % / 1.69% ± 0.92 %, respectively.

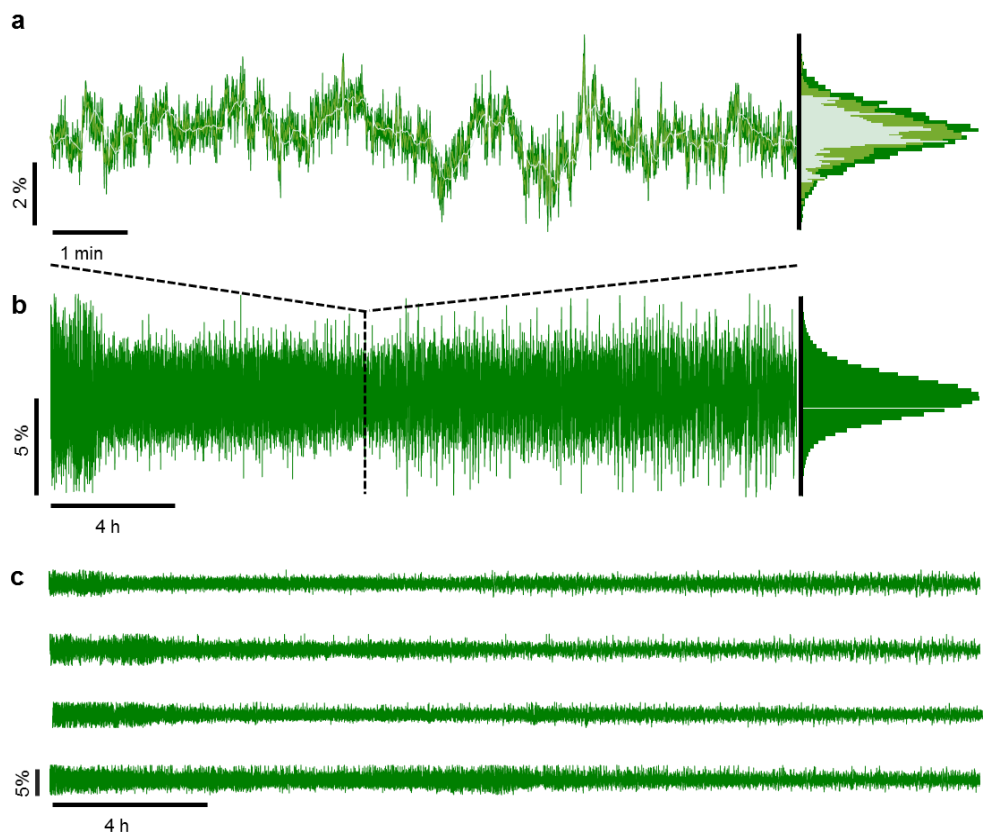

**Figure S6 | Time traces of Hsp90 connected plasmon rulers.** Time traces of plasmon rulers connected by Hsp90 molecules that are not fluctuating between the open/closed conformations. A zoom of 10 min is shown in **a**, the entire 24 hours measurement in **b**. **c**, Additional traces of different plasmon rulers. We evaluate the signal variation over 10 min. with the full time resolution (100ms, green line), with 1s time resolution (light green), and 10s time resolution (gray), c.f. histograms in **a**, right side. The observed signal variation of 2% is caused, in part, by the Hsp90 molecule.

# Examples of measurement resolution

In the above control experiments, we have determined the fluctuation amplitude measured with different time-resolution and over different periods of time. The results are summarized in **Table S1** below.

**Table S1** | Relative intensity fluctuations (standard deviation) measured at different time resolution/measurement time for the setup itself, the PEG linker and the Hsp90 molecule (without open/closed transition).

|  | 100ms over 10min | 1s over 10 min | 10s over 10 min |
|---|---|---|---|
| **Setup** | 0.28 ± 0.10 % | 0.21 ± 0.10 % | 0.17 ± 0.08 % |
| **PEG Linker** | 0.80 ± 0.62 % | 0.64 ± 0.58 % | 0.53 ± 0.49 % |
| **HSP90** | 1.98 ± 0.87 % | 1.85 ± 0.89 % | 1.69 ± 0.92 % |

We can convert the relative intensity fluctuation values in **Table S1** to interparticle distance changes $\Delta d = w \cdot \Delta I_{rel}$ using the linear conversion factor $w = 25$. These values give then the minimum distance change that can be resolved. We have summarized the result in the **Table S2** below.

**Table S2 | Interparticle distance resolution**

|  | 100ms over 10min | 1s over 10min | 10s over 10min |
|---|---|---|---|
| **Setup** | 0.88 ± 0.95 Å | 0.70 ± 0.83 Å | 0.58 ± 0.68 Å |
| **PEG Linker** | 2.00 ± 1.55 Å | 1.60 ± 1.45 Å | 1.33 ± 1.23 Å |
| **HSP90** | 4.95 ± 2.18 Å | 4.63 ± 2.23 Å | 4.23 ± 2.30 Å |

To measure conformational dynamics (as in this manuscript), the changes must be larger than those in the last row of **Table S2**. For example, to resolve conformational changes resulting in an interparticle distance change of $\Delta d = 5$ Å are possible to observe at a time resolution of 100 ms. We would like to point out, that this includes already real fluctuations by the molecule under investigation (Hsp90). These fluctuations within one conformational state contain information of the molecule itself.

# Limits of plasmon rulers

Our realization of the plasmon ruler technique, has not yet reached the maximum possible observation bandwidth. There should not be fundamental limit for the total observation time except for the lifetime of the molecule under investigation. On the other extreme, the data acquisition speed is fundamentally limited for two different reasons. Conformational dynamics faster than the gold sphere diffusion time cannot be measured. In our case, this time resolution would be 0.1 µs (calculated for a distance change $\Delta x$ of 1 nm and spheres of 60 nm diameter).

The other limit is the heating associated with the increase of illumination power necessary to yield an equivalent signal to noise ratio as in our measurements. If we tolerate a temperature increase of 1 K, we could reduce the time to about 7 µs (see below).

Technical improvements like a more efficient light collection, detectors with higher quantum yield, or lower read-out noise could reduce this time further. Both calculations indicate that with sufficiently fast detectors, protein conformational dynamics can possibly be observed with µs time resolution.

### Estimation of temperature increase

Our illumination ($I = 1.1 \cdot 10^4$ W/m$^2$) heats[9] the particles about $\Delta T = c_{abs} \cdot I / 4\pi\kappa r = 7 \cdot 10^{-4}$ K. In this estimation, we assume the particles to be surrounded by water with a thermal conductivity $\kappa = 0.591$ W/m·K and no significant barrier for the heat transfer from the particles to water. The light absorption cross section of our gold spheres is approximately $c_{abs} = 1.5 \cdot 10^{-14}$ m$^2$.

For a temperature increase of $\Delta T_{max} = 1$ K, we could therefore increase the illumination intensity by about $1/7 \cdot 10^{-4}$, resulting in a time resolution of about 7 µs.

## Additional time traces showing Hsp90 fluctuations qualitatively

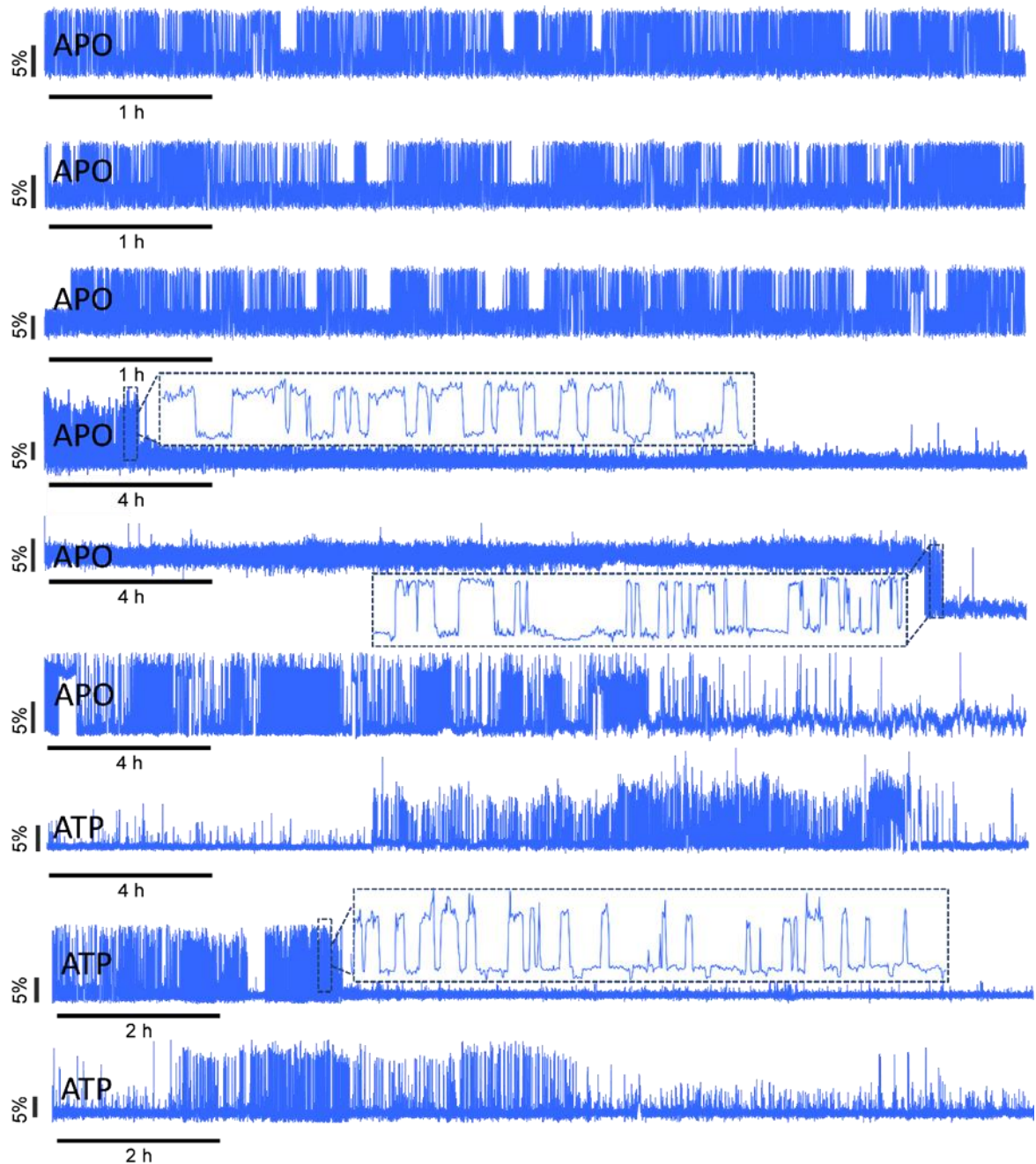

**Figure S7 | Additional timetraces showing Hsp90 fluctuations.** Timetraces of the dimers showing typical conformational transitions of Hsp90 complexes in the nucleotide-free protein (APO) and in the presence of ATP. The traces were recorded for 6, 12 and 24 hours with times resolutions between 10Hz to 50Hz.

## Decay rates

The table below shows the three decay rates for the open and closed state of Hsp90 obtained from a combination of all acquired traces. Under nucleotide-free condition (without ATP), we obtained 27700 dwell times, for the nucleotide containing buffer (with ATP), we observed 33000 transitions. We used a double-exponential decay function to extract the first two fast decay rates ($t_1$ and $t_2$) from the cumulative dwell time distribution $P(\tau)$.[1] The slow decay rate ($t_3$) was extracted by linear regression to the cumulative occurrence $- P(t >= \tau)$.

**Table S3** | Decay rates of sub-states within the open and closed configuration, determined from the dwell time distribution of plasmon ruler experiments in the presence and absence of ATP.

|  | Open | | | Closed | | |
|---|---|---|---|---|---|---|
|  | $t_1$ / s | $t_2$ / s | $t_3$ / s | $t_1$ / s | $t_2$ / s | $t_3$ / s |
| **ATP free** | 1.63 ± 0.03 | 13.2 ± 0.06 | 337 ± 70 | 0.78 ± 0.02 | 4.30 ± 0.11 | 115 ± 9 |
| **With ATP** | 0.39 ± 0.05 | 6.00 ± 0.48 | 250 ± 22 | 0.13 ± 0.02 | 2.10 ± 0.15 | 49.6 ± 2.0 |

## Hidden Markov Modelling

The plasmon ruler and FRET data were analyzed with a Hidden Markov Model (HMM) based approach.[10] The HMM yields the optimal kinetic model including the rates for all transitions. The rates were compared for HMMs trained with the plasmon and FRET data, respectively.

Overall, a good agreement of the rates is found for the transitions that occur most often in the data set (1→2 and 2→1). The different size of the confidence intervals (error bars) between plasmon and FRET data reflects the different size of the two data sets (864000 time points for plasmon and 25200 for FRET).

Even though the plasom data was recorded with a slightly higher sampling rate compared to the FRET data (10 Hz vs 5 Hz), transition rates are mostly smaller for the plasmon data. This reflects the fact that predominantly long dwells are missed with the FRET approach, thus overestimating the rate.

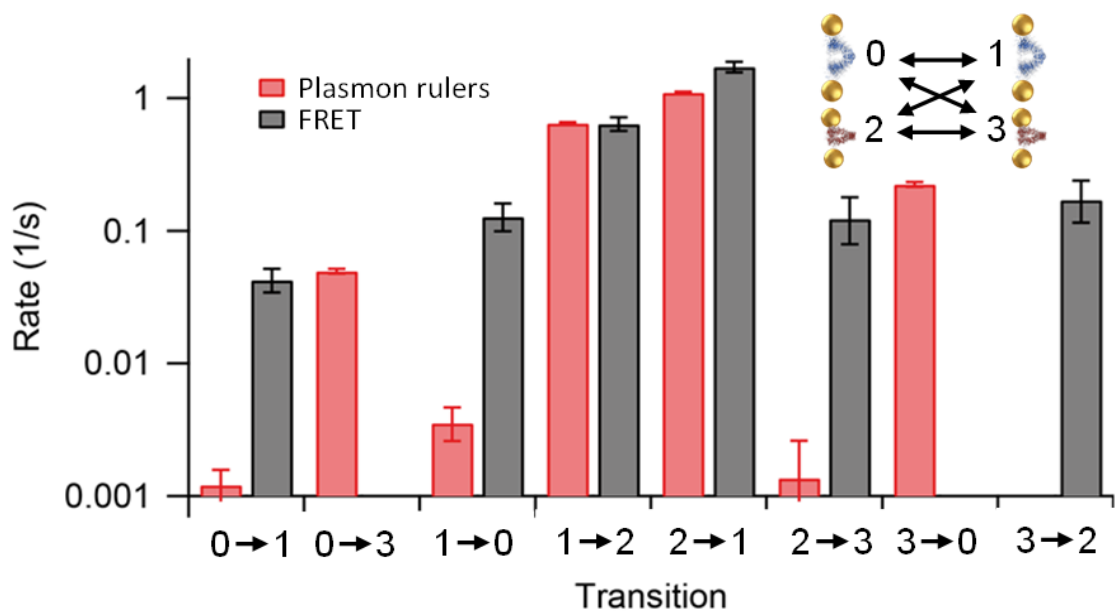

**Figure S8 | Rates from Hidden Markov Modelling.** The transition rates from a Hidden Markov analysis yields very similar rates between a FRET experiment (gray bars) and plasmon rulers (red bars).

# Variation between different single molecules

We compared the total times spend in the open/closed configuration for different molecules to test if the ergodic principle and the assumption of molecular homogeneity is correct. In **Figure S9**, we show the time (in %) the molecule spends in the open configuration for different molecules, both for buffers containing ATP (pink triangles) and lacking ATP (blue circles). The molecules in buffers containing ATP are mostly locked in the open position (90% of the time on average). However, there is a large variation between different molecules, showing values between 43% and 96%, which means that either the molecules are not as homogeneous as thought or the observation time of 12-24 hours is not enough to ensure ergodicity.

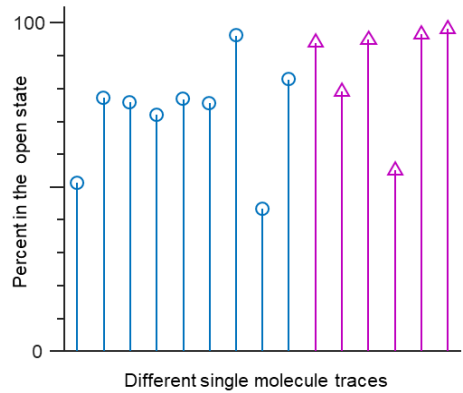

**Figure S9: Time spend in the open configuration for different molecules.** The molecules shown by blue lines and blue circles are measured without ATP in the buffer, the molecules shown by pink lines and triangles in an ATP containing buffer. The former molecules spend 75% of their time in open states on average, the latter 90% of the time. All traces are measured for 6 - 24 hours. The large deviations between molecules shows either non-ergodic behavior or molecular inhomogeneity.

# Signal to noise for single molecule FRET and plasmon rulers

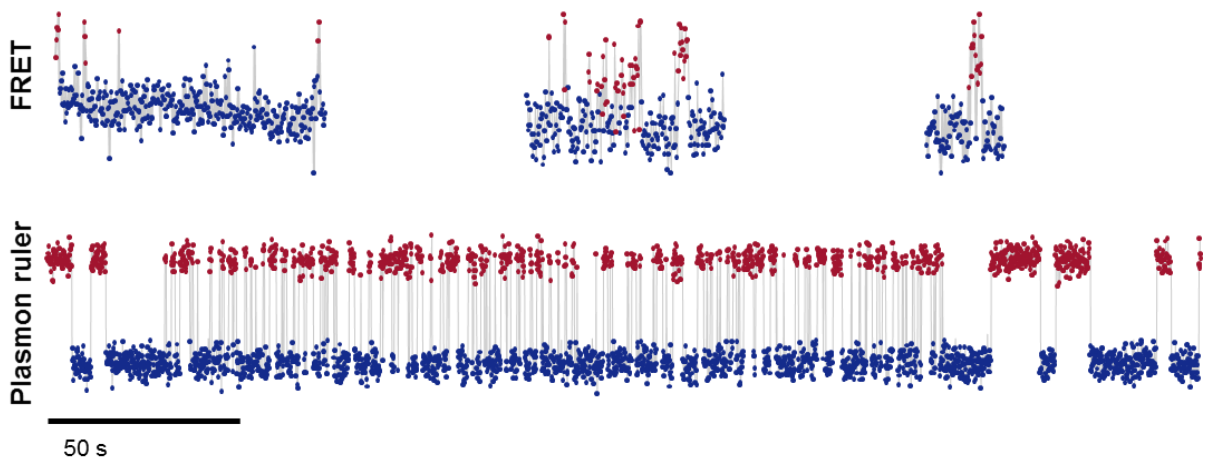

**Figure S10 | Comparison of single molecule traces from FRET and plasmon ruler measurements.** FRET traces (upper row) are significantly shorter and noisier than plasmon ruler traces (lower row). Both traces are shown with the same timescale. The FRET traces show the FRET efficiency which is a measure of the intermolecular distance – however, there is more information in the donor-/acceptor fluorescence (not shown here) that is used to assign the open/closed conformation. The plasmon ruler trace in the bottom shows only 300s out of 86400s (24 hours). The dots show the actual data points, colored according the open (blue) and closed (red) conformation, the thin gray lines connecting adjacent points is a guide to the eye.